\documentclass[twocolumn]{jpsj2} 
\def\vec#1{\boldsymbol{#1}}

\def\vr{\boldsymbol{r}}
\def\vq{\boldsymbol{q}}
\def\vk{\boldsymbol{k}}
\def\vPi{\boldsymbol{\it \Pi}}
\def\d{{\rm d}}
\def\e{{\rm e}}
\def\i{{\rm i}}

\title{Fulde--Ferrell--Larkin--Ovchinnikov State in Heavy Fermion Superconductors}

\author{Y. \textsc{Matsuda}$^{1,2}$ and H. \textsc{Shimahara}$^{3}$}

\inst{
$^{1}$Department of Physics, Kyoto University, Kitashirakawa, Sakyo-ku, Kyoto 606-8502, Japan\\
$^{2}$Institute for Solid State Physics, University of Tokyo, Kashiwanoha, Kashiwa, Chiba 277-8581, Japan\\
$^{3}$Department of Quantum Matter Science, ADSM, Hiroshima University, Higashi-Hiroshima 739-8530, Japan}

\abst{ The Fulde--Ferrell--Larkin--Ovchinnikov (FFLO) state is a novel superconducting state in a strong magnetic field characterized by the  formation of  Cooper pairs with nonzero total momentum  $(\vk\uparrow,  -\vk+\vq \downarrow)$, instead of the ordinary BCS pairs $(\vk \uparrow,  -\vk \downarrow)$.  A fascinating aspect of the FFLO state is that it exhibits inhomogeneous superconducting phases with a spatially oscillating order parameter and spin polarization. 
The FFLO state has been of interest in various research fields, not only in superconductors in solid state physics, but also in neutral Fermion superfluid of ultracold atomic gases and in color superconductivity in high energy physics. 
In spite of extensive studies of various superconductors, there has been no undisputed experimental verification of the FFLO state, mainly because of the very stringent conditions required of the superconducting materials.  Among several classes of materials, certain heavy fermion and organic superconductors are believed to provide conditions that are favorable to the formation of the FFLO state.   This review presents recent experimental and theoretical developments of the FFLO state mainly in heavy fermion superconductors.  In particular we address the recently discovered quasi-two-dimensional superconductor CeCoIn$_5$, which is a strong candidate for the formation of the FFLO state.}

\kword{
FFLO state, 
upper critical field, 
vortex state, 
heavy fermion superconductor, 
higher Landau level, 
phase diagram
}

\begin{document}
 \maketitle
\section{Introduction}

 Unconventional superconductivity with a nontrivial Cooper pairing state has been the subject of great theoretical and experimental interest.  Usually, unconventional superconductivity is characterized by anisotropic superconducting gap functions, which often have zeros (nodes) on the Fermi surface.   Realization of these nodal gap functions has been proved or strongly suggested for several classes of materials, including heavy-fermion, high-$T_{\rm c}$ cuprates, and organic superconductors.~\cite{Sig91,Min99,Tha05,Mat06} These unconventional superconductors are characterized by a superconducting order parameter which changes  sign in {\it momentum space} along  certain directions in the Brillouin zone.

A novel superconductivity that belongs to another class of unconventional superconducting state was predicted by Fulde and Ferrell~\cite{Ful64}, and Larkin and Ovchinnikov~\cite{Lar64} (FFLO) some 4 decades ago.   In the FFLO state, the superconducting order parameter changes sign in {\it real space}.  The FFLO state originates from the paramagnetism of conduction electrons. In the FFLO state, the Cooper pair formation occurs between  Zeeman splitted parts of the Fermi surface and a new pairing state $(\vk \uparrow,  - \vk + \vq \downarrow)$ with finite $\vq$ is realized; the Cooper pairs have finite center-of-mass momenta. This contrasts with $(\vk \uparrow,  - \vk \downarrow)$ pairing in the traditional BCS pairing state. Because of the finite $q \equiv\,|\vq|$, the superconducting order parameter 
$\Delta ( \vr ) \propto 
    \langle   \psi^{\dagger}_{\downarrow} ( \vr ) 
            \psi^{\dagger}_{\uparrow}   ( \vr ) \rangle $ 
has an oscillating component $\e^{\i \vq \cdot \vr} $ and a textured superconducting order parameter with a wave length of the order of the superconducting coherence length $\xi$ develops.  In the FFLO state, spatial symmetry breaking originating from the appearance of the $\vq$-vector appears, in addition to gauge symmetry breaking.  As a result, an inhomogeneous superconducting state appears.  Several fascinating superconducting properties which have never been observed for the BCS pairing state have been predicted for the FFLO state.~\cite{Cas04}

In spite of the straightforward nature of the theoretical prediction, this long sought for superconducting state has only been reported experimentally recently.   This is mainly because its occurance requires several very stringent conditions on the superconducting materials.   In most  type-II superconductors,  destruction of superconductivity occurs  through the orbital pair breaking effect,  leading to the emergence of the Abrikosov vortex state.   However, such an orbital effect is always detrimental to the FFLO state~\cite{Gru66}. Hence, for the FFLO state to occur, the orbital pair breaking effect must be weak relative to the Pauli paramagnetic effect, so that superconductivity survives up to the Pauli limit.  Moreover the system needs to be very clean, since the FFLO state is readily destroyed by impurities~\cite{Asl69,Tak70}.  Very few superconductors fulfill these necessary conditions for the FFLO state. Some of these conditions are satisfied in heavy fermion superconductors with very large orbital limiting fields and quasi-two-dimensional organic superconductors in a magnetic field applied nearly parallel to the conducting layers~\cite{Glo93,Yin93,Bur94,Shi94,Dup95,Tac96,Buz97}. In addition, anisotropies of the Fermi surface and superconducting gap function are favorable for stability of the FFLO state.

This review presents recent experimental and theoretical developments of the FFLO state, addressing the issue of its realization in heavy fermion superconductors.  In particular, we focus on the recently discovered quasi-two-dimensional superconductor CeCoIn$_5$, for which there is growing experimental evidence of an FFLO state.~\cite{Rad03,Bia03,Wat04,Wat04a,Cap04,Cor06,Kak05,Kum06,Mic06,Gra06,Mit06}

\section{FFLO state}

Generally, in type-II superconductors,  an applied magnetic field destroys the superconductivity in two distinct ways, orbital and paramagnetic pair-breaking effects.  The orbital effect leads to the emergence of an Abrikosov vortex state, forming a regular array of vortex lines parallel to the field.  Then, the kinetic energy of the superconducting currents around the vortex cores reduces the superconducting condensation energy.   The orbital limiting field is simply given as  $H_{\rm c2}^{\rm orb}=\Phi_0/2\pi\xi^2$  at which vortex cores begin to overlap.  Here $\Phi_0=\pi\hbar c/|e|$ is the flux quantum.   Experimentally, $H_{\rm c2}^{\rm orb}$ at $T=0$ is commonly derived from the slope of the determined $H$--$T$ phase boundary at $T_{\rm c}$ as \cite{Sai69}
\begin{equation}
     H_{\rm c2}^{\rm orb}(T=0) 
       = - 0.7 \,[{\rm K}] 
        \times  \left . \frac{\d H_{\rm c2}}{\d T} \right |_{T=T_{\rm c}}. 
\label{eq:Hc2orb}
\end{equation}

The paramagnetic pair-breaking effect originates from the Zeeman splitting of single electron energy levels.  When a magnetic field is applied in the normal state, electrons are polarized because the Fermi surface is split into spin-up and spin-down electron parts due to the Zeeman effect (Pauli paramagnetism). In contrast, in the superconducting state, spin-singlet Cooper pairs with vanishing spin susceptibility are not spin polarized.   Then, in order to polarize condensed electrons, the pairs must be broken.  This destruction of the superconductivity by  Pauli paramagnetic pair breaking occurs when the Pauli paramagnetic energy $E_{\rm P}=\frac{1}{2}\chi_{\rm n}H^2$ coincides with the superconducting condensation energy $E_{\rm c}=\frac{1}{2}N(0)\Delta^2$.  Here $\chi_{\rm n}=\frac{1}{2}(g\mu_{\rm B})^2N(0)$ is the spin susceptibility in the normal state, where $g$ is the spectroscopic splitting factor of an electron ($g = 2$ for a free electron),  $\mu_{\rm B}$ is the Bohr magneton,  $\Delta$ is the superconducting gap and $N(0)$ is the density of states at the Fermi level.  The Pauli limiting upper critical field is estimated to be $H_{\rm c2}^{\rm P}=\sqrt{2}\Delta/g\mu_{\rm B}$ (Chandrasekhar--Clogston limit)\cite{Cha62}.

Generally, the upper critical field is determined by  both the orbital and paramagnetic effects. The relative importance of the orbital and paramagnetic effects in the suppression of the superconductivity is described by the Maki parameter\cite{Sai69} 
\begin{equation}
  \alpha=\sqrt{2}\frac{H_{\rm c2}^{\rm orb}}{H_{\rm c2}^{\rm P}},
\label{eq:alpha}
\end{equation}
which is the ratio of  $H_{\rm c2}^{\rm orb}$ and $H_{\rm c2}^{\rm P}$ at zero temperature.  Since $\alpha$ is of the order of $\Delta/\varepsilon_{\rm F}$ , where $\varepsilon_{\rm F}$ is the Fermi energy,  $\alpha$ is usually much less than unity, indicating that the influence of the paramagnetic effect is negligibly small in most superconductors.  However, in materials with heavy electron effective mass, in which  the Fermi energy is small,  or in  layered materials in a magnetic field parallel to the layers, $\alpha$ can be even larger than unity.

\subsection{FFLO state in the pure Pauli limit}

We first discuss the superconducting phase of Pauli paramagnetically limited superconductors, in which the uniform magnetic field acts only on the spins of the conduction electrons and all orbital effects are neglected, corresponding to the limiting case of infinitely large Maki parameter.  It has been shown that the Pauli paramagnetic effect leads to some peculiar effects on the $H$--$T$ phase diagram. 

For simplicity, we discuss isotropic 2D and 3D superconductors with cylindrical and spherical Fermi surfaces, respectively, assuming $s$-wave superconducting gap symmetry. A 2D superconductor can be realized in a thin film in a magnetic field applied parallel to the plane of the superconductor.  When the film thickness is smaller than the coherence length,  the orbital pair-breaking effect is absent.

\begin{figure}[t]
\begin{center}
\includegraphics[width=7.5cm]{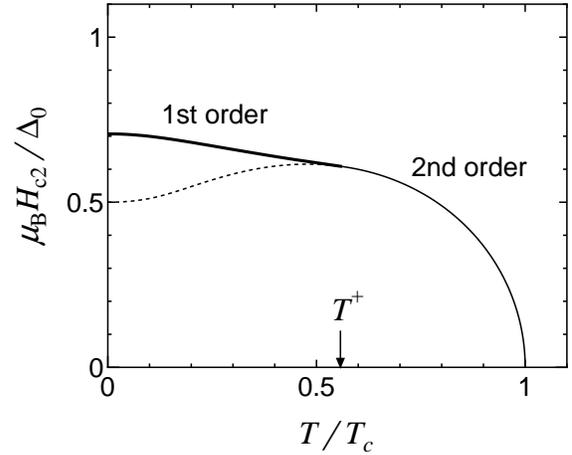}
\caption{$H$--$T$ phase diagram of 2D and 3D superconductors in the pure Pauli limit with no orbital effect.  $g=2$ is assumed.  For a 2D superconductor, $\vec{H}$ is applied parallel to the plane.   Below $T^{\dagger}~(=0.56T_{\rm c})$,  the transition becomes first order\cite{Sai69,Ket99}.  The broken line represents the metastable transition line. }
\label{fig:pauli1}
\end{center}
\end{figure}

Figure~\ref{fig:pauli1} displays the $H$--$T$ phase diagram of 2D and 3D superconductors in the Pauli limit with no orbital effect \cite{Sai69,Ket99}.  The initial slope of $H_{\rm c2}^{\rm P}$  at $T_{\rm c}$ is infinite and 
$\d H_{\rm c2}/\d T \propto \sqrt{T_{\rm c}/(T_{\rm c}-T)}$ 
in the vicinity of $T_{\rm c}$, in contrast to the orbital limited case with finite slope.  For the orbital limiting case with no Pauli effect, the transition at $H_{\rm c2}^{\rm orb}$ is of second order.  In contrast, for the Pauli limiting case, the upper critical field, assuming a second order transition line obtained by solving the gap equation, decreases with decreasing temperature below $T^{\dagger}\simeq 0.56T_{\rm c}$; {\it i.e.} the second order line at $T<T^{\dagger}$ represents a metastable state.  Then,  below $T^{\dagger}$,  the transition becomes first order at the upper critical field down to $T=0$.  The upper critical field is  $H_{\rm c2}^{\rm P}=\sqrt{2}\Delta/g\mu_{\rm B}$.

\begin{figure}[t]
\begin{center}
\includegraphics[width=7.5cm]{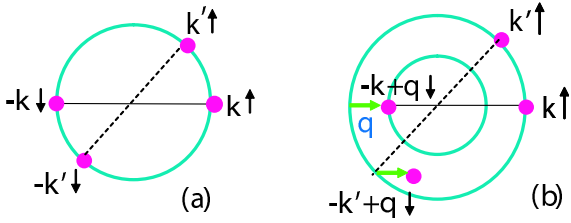}
\caption{Schematic figure of pairing states.  
(a) BCS pairing state $(\vk \uparrow,  -\vk + \vq \downarrow)$.  
(b) FFLO pairing state $(\vk \uparrow,  -\vk+\vq \downarrow)$.  
The inner and outer circles represent the Fermi surface of the spin down and up bands, respectively. The electron with $-\vk'+\vq \downarrow$ is not on the inner Fermi surface.}
\label{fig:pair}
\end{center}
\end{figure}

FFLO pointed out that the critical field can be further enhanced; the normal state is unstable with respect to the second order transition to an inhomogeneous superconducting state.  The Pauli paramagnetic pair-breaking effect is reduced by the formation of a new pairing state $(\vk \uparrow, -\vk + \vq \downarrow)$ with $q \sim 2 \mu_{\rm B} H / \hbar v_{\rm F}$, where $v_{\rm F}$ is the Fermi velocity, between the Zeeman splitted parts of the Fermi surface; spin-up and spin-down electrons can only stay bound if the Cooper pairs have finite center-of-mass momenta.  This contrasts with  \mbox{$(\vk\uparrow, -\vk\downarrow)$}-pairing in the BCS state. Figure~\ref{fig:pair} illustrates these pairing states.  Hereafter, we denote the BCS pairing state with $\vq = 0$ as the BCS state and the pairing state with $\vq \neq 0$ as the FFLO state. 

Compared to the BCS state, the FFLO state has a smaller condensation energy, butreduction of the Zeeman energy can stabilize the FFLO  state. Then upper critical field in the FFLO state $H_{\rm FFLO}$ can be higher than $H_{\rm c2}^{\rm P}$. At $T=0$,  $H^{\rm 3D}_{\rm FFLO} \simeq 1.51 \Delta_0 / g\mu_{\rm B} \simeq 1.07H_{\rm c2}^{\rm P}$ for 3D~\cite{Ful64,Lar64,Tak69,Sai69}, and $H^{\rm 2D}_{\rm FFLO} \simeq 2\Delta_0/g\mu_{\rm B} \simeq 1.42H_{\rm c2}^{\rm P}$ for 2D.~\cite{Bul73,Aoi74,Bur94,Shi94}  For 1D, $H^{\rm 1D}_{\rm FFLO}$ diverges as $T \rightarrow 0$, as will be discussed later.~\cite{Suz83,Mac84}  It turns out that the FFLO state only appears at \mbox{$T<T^{\dagger}$}; the point $(T^{\dagger}, H^{\dagger})$ is a tricritical point, at which the normal, uniform BCS superconducting, and FFLO phases all meet.

\begin{figure}[t]
\begin{center}
\includegraphics[width=7.5cm]{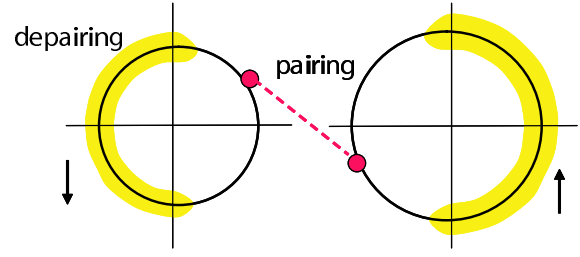}
\caption{Pairing state described by eq.~(\ref{eq:ff}).  The shaded region is the blocking region where Cooper pairs are not formed.}
\label{fig:ff}
\end{center}
\end{figure}

The  $\vq$-vector in the FFLO state gives rise to  spatial symmetry breaking. As a result, the superconducting order parameter exhibits spatial oscillation. A possible superconducting order parameter, originally proposed by Fulde and Ferrell, is 
\begin{equation}
     \Delta(\vr) = \Delta_1 \e^{\i \vq \cdot \vr},
\label{eq:ff}
\end{equation}
in which the amplitude of the superconducting order parameter is homogeneous, but the phase changes in real space. In this state,  depairing occurs for some part of the Fermi surface (the shaded portion in Fig.~\ref{fig:ff}) and pairing can ocur in the remaining regions of the Fermi surface.\cite{Ful64,Tak69}

In the gap equation, the two solutions $\Delta_1 \e^{\i \vq \cdot \vr}$ and $\Delta_1 \e^{- \i \vq \cdot \vr}$ represent degenerate superconducting states at a transition for a given $\vq$.  The degeneracy is lifted by forming the linear combinations of $\e^{ \i \vq \cdot \vr}$ and $\e^{ - \i \vq \cdot \vr}$. It has been found that 
\begin{equation}
     \Delta( \vr ) = \Delta_1 ( \e^{\i \vq \cdot \vr} + \e^{-\i \vq \cdot \vr})
                   = 2 \Delta_1 \cos( \vq \cdot \vr )
\label{eq:fflo1d}
\end{equation}
provides a lower free energy than eq.~(\ref{eq:ff}) for the whole temperature range below $T^{\dagger}$.~\cite{Lar64} This state was originally proposed by Larkin and Ovchinnikov. In this state, the superconductor becomes spatially inhomogeneous, because the order parameter undergoes 1D spatial oscillation with a wave length of  $2 \pi/q$ and, as a result, the normal quasiparticle and  paramagnetic moments appear periodically.

Generally, because of the symmetry of the system, there are more than two equivalent $\vq$-vectors which give the same upper critical field.  Then, the order parameter is expressed by a linear combination of $\e^{\i \vq_m \cdot \vr}$, 
\begin{equation}
     \Delta( \vr ) = \sum_{m=1}^{M} \Delta_m \e^{\i \vq_m \cdot \vr}, 
\end{equation}
where $M$ denotes the number of equivalent $\vq$-vectors.~\cite{Lar64,Shi98a,Bow02,Mor04,Mor05,Wan06,Com05} In an isotropic system, $M = \infty$.

\begin{figure}[t]
\begin{center}
\includegraphics[width=7.5cm]{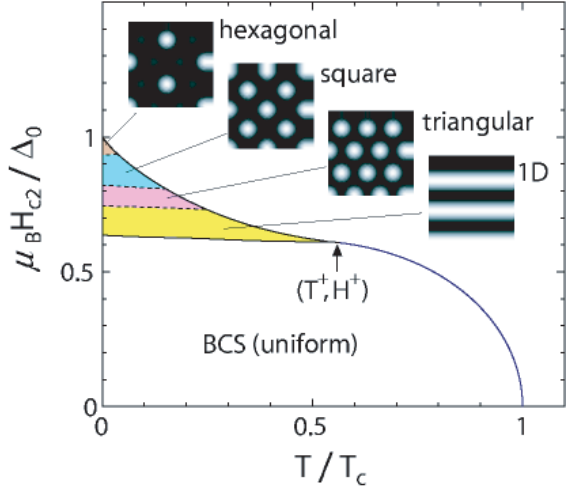}
\caption{$H$--$T$ phase diagram of a 2D superconductor in a parallel field. The color regions are in the FFLO state.  $(T^{\dagger}, H^{\dagger})$ is the tricritical point. The solid  and dashed lines represent the second and first order transition lines, respectively.  Just below $T^{\dagger}$, the order parameter undergoes 1D spatial oscillation, as expressed by eq.~(\ref{eq:fflo1d}).  At lower temperatures, triangular, square, and hexagonal FFLO states, expressed by eqs.~(\ref{eq:fflotri}), (\ref{eq:fflosqu}) and (\ref{eq:fflohex}), respectively, appear.  For details, see the text.  The quasiparticle densities of the states in each FFLO state are shown.  }
\label{fig:fflo2d}
\end{center}
\end{figure}

Figure~\ref{fig:fflo2d} displays the $H$--$T$ phase diagram and quasiparticle structure in a 2D thin film in a parallel field. Just below $T^{\dagger}$, the order parameter described by eq.~(\ref{eq:fflo1d}) is stabilized.  However, as the temperature decreases, order parameters with linear combinations of larger number of plane waves stabilize~\cite{Shi98a}. This is because the increasing number of plane waves increases the number of nodes, which in turn reduces the Pauli paramagnetic energy of the quasiparticles excited around the nodes.  At lower temperatures, the ``triangular state'',  with a gap function of the form 
\begin{equation}
\Delta( \vr ) = \Delta_1 \, (  \e^{\i \vq_1 \cdot \vr} 
                             + \e^{\i \vq_2 \cdot \vr} 
                             + \e^{\i \vq_3 \cdot \vr} ) ,
\label{eq:fflotri}
\end{equation}
has a lower free energy, 
with $\vq_1 = (q,0,0) $, 
     $\vq_2 = (-q/2,\sqrt{3}q/2,0) $, 
 and $\vq_3 = (-q/2,-\sqrt{3}q/2,0) $, 
where we take the $z$-axis to be perpendicular to the 2D plane. 
A further decrease of the temperature stabilizes the ``square state'', with a gap function of the form 
\begin{equation}
     \Delta(\vr) = \Delta_1 \, (\cos (qx) + \cos(qy)), 
\label{eq:fflosqu}
\end{equation}
and the ``hexagonal state'',  with a gap function of the form 
\begin{equation}
\Delta(\vr) = \Delta_1 \, (  \cos( \vq_1 \cdot \vr ) 
                               + \cos( \vq_2 \cdot \vr ) 
                               + \cos( \vq_3 \cdot \vr ) ). 
\label{eq:fflohex}
\end{equation}

In a 3D isotropic system,  a recent calculation~\cite{Mor05} demonstrates that the first order transition line is slightly higher than the second order transition line at all temperatures below $T^{\dagger}$.  Moreover, similar to the 2D case, the order parameter is $\cos(\vq\cdot \vr)$  just below $T^{\dagger}$, but at lower temperatures the order parameter switches to more complicated structures which are expressed by a sum of two or three cosines.  It has also been suggested that at zero temperature a state, which is expressed by a sum of eight plane waves forming a face-centered cubic structure, appears.~\cite{Bow02}

We note that the FFLO state had not been reported for thin films until now.  The main reason for this is that the FFLO state is readily destroyed by impurities,~\cite{Asl69,Tak70} as will be discussed later.  In fact, most thin films are in the dirty limit, $\xi \gg \ell$, where $\ell$ is the mean free path of the electrons.  Even in clean films, surface scattering may destroy the FFLO state. Even if a very clean thin film could be produced,  the magnetic field must be applied very precisely parallel to the plane  to produce the FFLO state.~\cite{Shi97b}  Otherwise a small but finite perpendicular field component leads to the formation of a vortex state, which destroys the FFLO state (see \S2.2).

The stability of the FFLO state with respect to thermal fluctuations is different from that of BCS superconductors~\cite{Shi98b,Oha02}. In an isotropic system, the direction of the $\vq$-vector is arbitrary.  For 3D systems, long-range-order (LRO) is destroyed by thermal fluctuations of the $\vq$-vector and  quasi long range order (QLRO),  characterized by a correlation function with a power law decay, is established~\cite{Shi98b,Oha02}. For 2D systems, it is well known that  LRO is absent even in the BCS state, and superconductivity with QLRO is achieved through a Kosterlitz--Thouless transition.   In the FFLO state with 1D oscillation expressed by eqs.~(\ref{eq:ff}) and (\ref{eq:fflo1d}), even QLRO is absent due to  directional fluctuation of the $\vq$-vector.~\cite{Shi98b,OSC}  However, it has been shown that  LRO is established even in an FFLO state in the presence of crystal anisotropy~\cite{Shi98b}, which exists in real systems, because the anisotropy fixes the $\vq$-vector in a certain direction. It has also been shown that the fluctuations lead to a divergent spin susceptibility and a breakdown of the Fermi-liquid behavior at the quantum critical point~\cite{Sam06}.  Moreover, in the FFLO state of quasi-2D $d$-wave superconductors, a fluctuation-driven first-order transition has been proposed.~\cite{Dal04}

\subsection{Orbital pair-breaking}

In real bulk type-II superconductors, the orbital effect is invariably present.  Gruenberg and Gunther examined the stability of the FFLO state against the orbital effect in 3D isotropic systems with $s$-wave pairing.~\cite{Gru66}  It has been shown that  orbital effects are detrimental to the formation of the  FFLO state.  The FFLO state can exist at finite temperatures if $\alpha$ is larger than 1.8. The FFLO region shrinks considerably from that in the absence of the orbital effect.

In the presence of a magnetic field $\vec{H} = (0,0,H)$,  the spatial variation of the order parameter is determined by the cyclotron motion of the Cooper pairs in the plane perpendicular to $\vec{H}$. Generally, the orbital effect forces the solutions for the order parameter to be eigenfunctions of the operator $\vPi^2$, where $\boldsymbol{\vPi} = - {\rm i} \hbar {\boldsymbol \nabla} - \frac{2e}{c} \vec{A}$ with the vector potential $\vec{A} = (0,Hx,0)$. This problem is equivalent to that of a charged particle in a constant magnetic field. 
The operator $\vPi^2$ can be rewritten as 
\begin{equation}
     \vPi^2 = \frac{2\hbar^2}{\xi_H^2} \bigl ( \eta^{\dagger} \eta + \frac{1}{2} \bigr ) 
             - \hbar^2 \frac{\partial^2 }{\partial z^2} , 
\end{equation}
where $\eta^{\dagger}$ and $\eta$ are 
the boson creation and annihilation operators defined by 
\begin{equation}
\label{eq:etadef}
    \begin{array}{rcl}
      \eta           
      \hspace{-2ex} & = & \hspace{-2ex}
      \displaystyle{ \frac{\xi_H}{\sqrt{2}\hbar} (\Pi_x - {\rm i} \Pi_y) }\\[8pt]
      \eta^{\dagger} 
      \hspace{-2ex} & = & \hspace{-2ex}
        \displaystyle{ \frac{\xi_H}{\sqrt{2}\hbar} (\Pi_x + {\rm i} \Pi_y) } , 
    \end{array}
\end{equation}
respectively. Here, $\xi_H = \sqrt{\Phi_0/2\pi H}$. Eigenfunctions with respect to the degrees of freedom perpendicular to $\vec{H}$ are described by the Abrikosov functions with Landau level index $n$,
\begin{equation}
     \phi_n^{(k)}(\vec{\rho}) = (-1)^n  \e^{\i k y} 
        H_n \bigl[ \sqrt{2} \frac{x-x_k}{\xi_H} \bigr ] , 
\end{equation}
for any real number $k$ and  $x_k=k\xi_H$, where $H_n$ is the Hermite polynomial and  $\vec{\rho} = (x,y)$.  The lowest Landau level solution ($n=0$) is given by
\begin{equation}
     \phi_0^{(k)}(\vec{\rho}) = \e^{\i ky}
       \exp \bigl [ -\frac{(x-x_k)^2}{2\xi_H^2} \bigr ] .
\end{equation}

In the BCS state, the eigenvalues of the operator $\vPi^2$ are 
\begin{equation}
     \frac{2\hbar^2}{\xi_H^2} \bigl ( n + \frac{1}{2} \bigr ) . 
\end{equation}
Usually, the BCS state is described by the lowest Landau level solution
\begin{equation}
     \Delta(\vr) \propto \phi_0^{(k)} (\Vec{\rho}) , 
\end{equation}
which gives the highest upper critical field.

The coexistence of both the FFLO state and vortex modulation has been investigated through exact calculations of the upper critical field assuming that the transition is of second order\cite{Gru66,Tac96,Shi96a}.  In the FFLO state, the eigenvalues of the operator $\vPi^2$ are 
\begin{equation} 
     \frac{2\hbar^2}{\xi_H^2} \bigl ( n + \frac{1}{2} \bigr ) + \hbar^2 q^2.
\label{eq:kin} 
\end{equation}
The second term represents the kinetic energy of the Cooper pairs along $\vec{H}$.   We note that the increase of the kinetic energy in eq.~(\ref{eq:kin}) is compensated by the reduction of the Pauli paramagnetic energy.  Assuming that the lowest Landau level gives the highest upper critical field even in the FFLO state,  the order parameter is given by
\begin{equation}
     \Delta(\vr) \propto \exp(\i qz) \, \phi_0^{(k)}(\vec{\rho})
\label{eq:expphi}
\end{equation}
or
\begin{equation}
     \Delta(\vr) \propto \cos(qz) \, \phi_0^{(k)}(\vec{\rho}).
\label{eq:cosphi}
\end{equation} 
As indicated by eq.~(\ref{eq:cosphi}),  the order parameter exhibits a modulation with wave vector $\vq$ parallel to $\vec{H}$ and 2D nodal sheets appear perpendicular to $\vec{H}$. Equation~(\ref{eq:cosphi}) clearly indicates that  parallel motions of the Cooper pairs create the 2D nodal sheets.

\begin{figure}[t]
\begin{center}
\includegraphics[width=7.5cm]{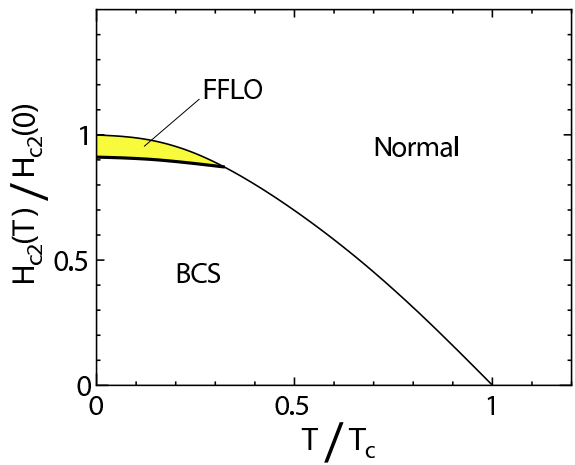}
\caption{$H$--$T$ phase diagram in the presence of the orbital effect, as proposed in Ref.~[\citen{Gru66}]  The transitions from the normal state to the FFLO and BCS states are of second order (solid line), while the transition from the FFLO state to the BCS state is of first order (thick solid line).}
\label{fig:gg}
\end{center}
\end{figure}

\begin{figure}[t]
\begin{center}
\includegraphics[width=7.0cm]{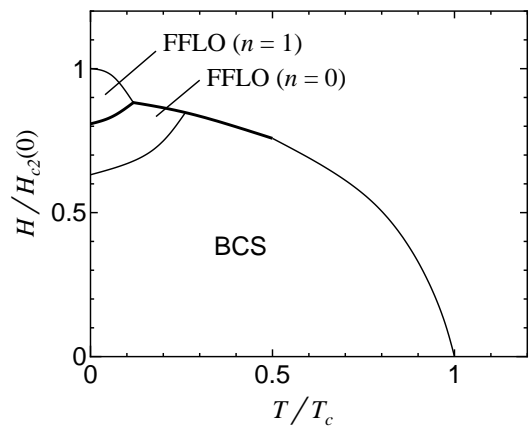}
\caption{$H$--$T$ phase diagram for a 3D system with large Maki parameter ($\alpha=1$)~\cite{Ada03}. $n$ denotes the Landau level index.  The thick and thin solid lines represent the first and second order transition lines, respectively.}
\label{fig:adachi}
\end{center}
\end{figure}

The $H$--$T$ phase diagram in the presence of the orbital effect proposed by Gruenberg and Gunther is schematically shown in Fig.~\ref{fig:gg}.  When the orbital effect is not weak, the transitions from the normal state to the FFLO and BCS states are of second order, while  the transition from the FFLO to the BCS state is of first order. Recent calculations indicate that the phase diagram is modified in some cases.\cite{Shi96a,Ada03,Hou06}   For example,  according to Ref.~[\citen{Ada03}],  in which the 4th and 6th order terms of the Ginzburg--Landau free energy are incorporated, some portion of the transition line at the upper critical field is of first order, while the transition from the FFLO state to the BCS state that branches from the first order line is of second order, as shown in Fig.~\ref{fig:adachi}.

\begin{figure}[b]
\begin{center}
\includegraphics[width=7.0cm]{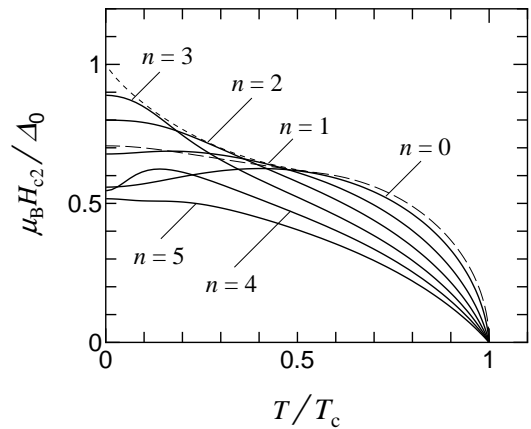}
\caption{Upper critical fields of states with $n$ of a 2D system in a magnetic field slightly tilted from the plane of the superconductor\cite{Shi97b}. 
$n$ denotes the Landau level index. 
The transitions at $H_{\rm c2}$ are assumed to be of second order. 
The physical critical field is the maximum one at each temperature. 
The broken and dotted lines show the critical fields of the BCS state ($\vq = 0$) and the FFLO state ($\vq \neq 0$) in a parallel field, respectively. }
\label{fig:2Dtilt}
\end{center}
\end{figure}

A further increase in the relative importance of the Pauli paramagnetic effect produces an FFLO state with the Abrikosov functions of a higher Landau level~\cite{Bul73,Shi97b,Shi96a,Buz96,Ada03}.  This is because the Pauli paramagnetic energy is reduced as the number of nodes is increased,  which in turn increases the number of polarized quasiparticles in the vicinity of the nodes.  This situation would be realized in heavy fermion superconductors with very large Maki parameter and in 2D layered superconductors in magnetic fields slightly tilted from the 2D plane.  
In a 2D system, very small perpendicular component of $\vec{H}$ leads to a very large effective Maki parameter~\cite{Bul73,Shi97b,Buz96}. Figure~\ref{fig:2Dtilt} shows a phase diagram of a 2D system in a slightly tilted magnetic field,  obtained in Ref.~[\citen{Shi97b}]. A similar phase diagram has also been obtained in Ref.~[\citen{Buz96}].  A rich variety of quasiparticle structures associated with the higher Landau level indexes appear as a function of the tilted angle and temperature~\cite{Kle00,Hou00,Kle04,Yan04}. 
For a 3D system, it has been shown that an FFLO state can coexist with the Abrikosov state with higher Landau levels~\cite{Shi96a}.  Figure~\ref{fig:adachi} schematically shows the $H$--$T$ phase diagram for a 3D system with large Maki parameter as proposed in Ref.~[\citen{Ada03}]. At very low temperatures, the Abrikosov function with the first Landau level ($n=1$) provides the highest upper critical field. In this state, nodal planes are formed periodically parallel to the vortex lattice.

\subsection{Anisotropies of Fermi surface and gap function}

As discussed in the previous sections, the FFLO state is restricted in an extremely narrow region in the $H$--$T$ plane ($H_{\rm FFLO}^{\rm 3D} \approx 1.07 \times H_{\rm c2}^{\rm P}$ in the absence of the orbital pair-breaking effect). In this circumstance,  the FFLO state is readily destroyed by several effects, such as internal field enhancement due to electron correlations (Fermi liquid correction)~\cite{Tak69,Bur94,Shi94,Vor06}. In fact, a very slight enhancement of the uniform susceptibility $\chi$ from that without correlations $\chi_0$, $\chi/\chi_0 = 1.04$, is large enough to suppress the FFLO state~\cite{enhance}. For stability of the FFLO state in real systems, it is necessary that the Fermi surface or the gap function be anisotropic. These anisotropies stabilize the FFLO state through optimization of the $\vq$-vector.

The stability of the FFLO state also crucially depends on the dimensionality of the system.  This can be understood from Fig.~\ref{fig:pair}. The FFLO state is more stable when a large area of the spin-up Fermi surface is connected to the spin-down surface by the $\vq$-vector. This bears some analogy with the nesting effect of charge-density-wave (CDW) and spin-density-wave (SDW) transitions~\cite{Shi94,Shi97a}.  For a 3D system with a spherical Fermi surface,  the spin-down and spin-down Fermi surfaces touch only at a point by a shift of the $\vq$-vector.  For a 2D system with a cylindrical Fermi surface,  the two Fermi surfaces touch on a line by a shift of the $\vq$-vector.  A particular case is a 1D system with a flat Fermi surface where the whole region of the Fermi surfaces touch by a single $\vq$-vector.  Therefore, low dimensionality favors the FFLO state, extending the FFLO regime.  However, for the 1D case, CDW or SDW transitions would occur at a finite temperature and the Fermi surface disappears.  Therefore, the best system to observe the FFLO state appears to be a 2D system.~\cite{Shi94}  Even in a 3D system, if the Fermi surface has some flat regions, the FFLO state is expected to be stabilized when $\vec{H}$ is oriented in an optimal direction of the $\vq$-vector.

The above argument considers the FFLO state for superconductors with isotropic $s$-wave symmetry.  Most of the candidate compounds with large Maki parameter belong to two classes of materials, quasi-2D organic and heavy fermion superconductors.  All these materials have been argued to be unconventional superconductors with anisotropic gap functions, most likely to have nodes, and in this way differ from the case originally considered by FFLO.  Hence, the experimental identification of nodal structures has stimulated theoretical interest in the FFLO state in anisotropic $d$-wave pairing symmetry~\cite{Mak96,Shi96a,Shi97a,Shi97b,Yan98,Vor05a}.

It has been shown that the stability of the FFLO state also depends on gap anisotropy. The gap function describing the FFLO state of anisotropic pairing has the form, for example, 
\begin{equation}
     \Delta(\Vec{r},{\Vec{\hat k}}) 
          \propto 
               \gamma({\Vec{\hat k}}) 
               \cos(qz) 
               \sum_{n=0}^{\infty} C_n 
               \phi_n^{(k)}(\Vec{\rho}),
\end{equation}
with ${\Vec{\hat k}} \equiv \Vec{k}/k$, where $\vr = (\vec{\rho},z)$ is the center-of-mass coordinate and $\vk$ is the relative momentum of the Cooper pair~\cite{Shi97b,Shi96a}. Here, $\gamma({\Vec{\hat k}})$ describes the momentum dependence of anisotropic pairing.  For $d_{x^2-y^2}$-wave pairing, $\gamma({\Vec{\hat k}}) \propto {\hat k}_x^2 - {\hat k}_y^2$, and for $d_{xy}$, $\gamma({\Vec{\hat k}}) \propto {\hat k}_x{\hat k}_y$. For $d$-wave symmetry in a 2D system with a cylindrical Fermi surface, the FFLO state is more stable when the $\vq$-vector is oriented in the direction of maximum gap at low temperatures. In actual systems, the optimal $\vq$-vector depends on both the Fermi surface and gap function anisotropies~\cite{Shi97a,Ike04}. In $\kappa$-(BEDT-TTF)$_2$X~\cite{Shi97a}, 
the FFLO state has been studied by taking into account both effects.

\subsection{Impurity Effect}

Impurity effects on the FFLO state have been studied by several authors~\cite{Asl69,Tak70,Hou06,Ada03}.  It has been shown that the FFLO state is unstable against nonmagnetic impurities, in contrast to the BCS state of $s$-wave pairing.  This is analogous with anisotropic superconductivity, in which the mixing of order parameters with opposite signs in momentum space caused by impurity scatterings suppresses the pair amplitude.  Similarly, in the FFLO state, mixing by impurity scattering in real space suppresses the pair amplitude.

It has also been shown that non-magnetic impurities dramatically change the  FFLO phase diagram.  In the presence of impurities, the behavior of the FFLO phases in unconventional superconductors is qualitatively different from that in $s$-wave superconductors~\cite{Agt01}.  For example, it has been suggested that in $d$-wave superconductors,  states with order parameters expressed by eqs.~({\ref{eq:ff}}) and ({\ref{eq:fflo1d}}) both appear in the same $H$--$T$ plane.

A possible change of the order of the phase transition from normal to the BCS and FFLO states due to impurities has also been suggested. A rich variety of phase diagrams has been proposed when both the orbital pair-breaking  and impurity effects are taken into account.~\cite{Hou06}

\subsection{Quasiparticle structure of a vortex in the FFLO state}

\begin{figure}[t]
\begin{center}
\includegraphics[width=8.5cm]{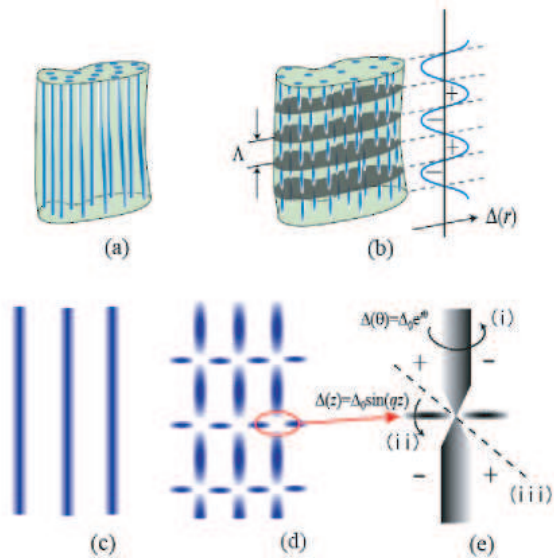}
\caption{(a) Schematic view of the flux line lattice in the BCS state.  (b) Flux line lattice in the FFLO state. In the FFLO state, planar nodes appear periodically perpendicular to the flux line, which leads to a segmentation of the vortices into pieces of a length  $\Lambda=2\pi/q$. (c) Schematic figure of the quasiparticle excitation spectrum in the BCS state.  The quasiparticles are excited within the vortex cores (blue region).  (d) Quasiparticle excitation spectrum in the FFLO state. The quasiparticles here are also excited around the FFLO planar nodes.    The quasiparticle regions within the vortex lines do not spatially overlap with those in the FFLO planar nodes. (e) Schematic view of (vertical) the vortex line and (horizontal) the FFLO nodal plane. Quasiparticles experience sign change of $\Delta(\vr)$ when tracing either path (i) or (ii) through the $\pi$ shift of the phase in the order parameter, while no sign change occurs when tracing a path  (iii).  For details, see the text. }
\label{fig:qpexc}
\end{center}
\end{figure}

The quasiparticle structure in the superconducting state in a magnetic field has been studied extensively.  In the BCS state, the quasiparticles are spatially localized within vortex cores forming Caroli--de Gennes--Matricon bound states with discrete energy levels\cite{Car64}.  The quasiparticle structure in the Pauli limit without the orbital effect has been discussed in \S 2.1. (For the 2D case, see Fig.~\ref{fig:fflo2d}.)  The quasiparticle structure in the  FFLO state in the presence of the orbital effect has been addressed by several authors.~\cite{Tac96,Miz05a,Miz05b}  Recently, the low energy quasiparticle excitation and spin polarization spectrum in the FFLO state for the isotropic 3D case has been examined in the framework of the microscopic Bogoliubov--de Gennes theory, taking into account the orbital and  Pauli paramagnetic effects in an equal footing.~\cite{Miz05a}  Figures~\ref{fig:qpexc}(c) and \ref{fig:qpexc}(d) display schematically the quasiparticle excitations in the BCS  and  FFLO states, respectively.  It is remarkable in the FFLO state that the quasiparticles are not excited in the region where the (vertical) vortex lines intersect the (horizontal) planar nodes.

This peculiar spatial quasiparticle structure is characteristic of  the topology of the FFLO vortex, as schematically shown in Fig.~\ref{fig:qpexc}(e).  Generally, quasiparticles are spatially localized  at the region where the phase of the superconducting order parameter is shifted by $\pm \pi$\cite{Tak80}; the bound states appear as a result of interference of the superconducting wave function caused by Andreev reflection by pair potentials with opposite signs. In Fig.~\ref{fig:qpexc}(e), (i) and (ii) represent the path cutting through the vortex line with 2$\pi$-phase winding ($\Delta \propto \e^{\i \theta}$) and the path crossing the FFLO nodal plane ($\Delta \propto \sin(\vq \cdot \vr)$, where we take the FFLO nodal plane to be $z=0$, respectively.  Quasiparticles tracing either path (i) or (ii) experience the sign change of $\Delta(\vr)$ through the $\pi$ shift of the phase in the order parameter.  The phase of the order parameter around the vortex line is additionally twisted by $\pi$ when crossing the FFLO sheet.  When tracing the path (iii) (the dotted line in Fig.~\ref{fig:qpexc}(e)), the quasiparticles do not experience a sign change of the pair potential, because of a $ \pi \pm \pi$ phase shift.  Therefore, bound states are not formed in the region where the vortex line intersects with the FFLO sheet.

The peculiar quasiparticle structure in the FFLO state is expected to change the  elastic properties of the flux line lattice (FLL)~\cite{Tac96,Wat04}. Notably, the occurrence of  FFLO nodal sheets perpendicular to the vortex lattice is expected to lead to a softening of the FLL tilt modulus $c_{44}$. This is due to the fact that a deformation of the nodal planes entails a deformation of the FLL (see \S 5.3). The decreased order parameter in the FFLO state  reduces the pinning strength that characterizes the FLL interaction with crystal defects. 
In other words, defects in the vicinity of the nodal planes no longer interact with the FLL, even though pinning of the nodal planes is  possibile. Defects between the nodal planes exert a smaller pinning force (proportional to $|\Delta|^{2}$).

\section{Candidate compounds for the  FFLO state}

Summarizing the  requirements for the formation of the FFLO state: 
\begin{enumerate}
\item Strongly type--II superconductors with very large Ginzburg--Landau parameter $\kappa \equiv  \lambda/\xi \gg 1$ and large Maki parameter $\alpha$, 
such that the upper critical field can easily approach the Pauli paramagnetic limit.
\item Very clean, $\xi \ll \ell$, since the FFLO state is readily destroyed by impurities.
\item Anisotropies of the Fermi--surface and the gap function can stabilize the FFLO state. 
\end{enumerate}
The main reason that the FFLO state has never been observed in conventional  superconductors resides in the fact that the requirements for the formation of the FFLO state are extremely hard to fulfil.  The question of observing the FFLO state in unconventional superconductors has been addressed recently.  Heavy fermion and organic superconductors are attractive classes of superconductors for forming the FFLO state because of their potentially large  Maki parameters, very large mean free paths, and  unconventional pairing symmetries.~\cite{Mat06}

Heavy electron masses lead to low Fermi velocities of the quasiparticles and, in turn, enhance  $H_{\rm c2}^{\rm orb}$ and the Maki parameter.  For this reason, heavy fermion superconductors have had considerable attention.  In fact, the features of the magnetization and the $H$--$T$ phase diagrams of  CeRu$_2$\cite{Mod96,Yam97}, UPd$_2$Al$_3$,\cite{Glo93,Mod96,Tac96} and UBe$_{13}$ were taken as possible signatures of the FFLO state.  Very recently, a very strong candidate for exhibiting the FFLO state,  CeCoIn$_5$, has been discovered for which a new superconducting phase has been confirmed by the thermodynamic second order transition.

In layered organic superconductors, when a magnetic field is applied parallel to the layers, the orbital effect is strongly suppressed.~\cite{Shi97a,Ish00} 
Indications for a possible FFLO state have been reported for
$\lambda$-(BETS)$_2$GaCl$_4$,~\cite{Tan02} 
$\lambda$-(BETS)$_2$FeCl$_4$,~\cite{Uji01,Bal01,Shi02a,Hou02,Man00,Uji06} 
and $\kappa$-(BEDT-TTF)$_2$Cu(NCS)$_2$.~\cite{Nam99,Sin00,Shi97a,Iza02a} 
For $\lambda$-(BETS)$_2$GaCl$_4$, a kink in the thermal conductivity suggests a transition  to the FFLO state.~\cite{Tan02} For $\kappa$-(BEDT-TTF)$_2$Cu(NCS)$_2$ a similar feature in the magnetization has been identified.~\cite{Nam99,Sin00} In spite of several observations which appear to support the FFLO state,  evidence of the thermodynamic phase transition between the FFLO and the BCS-pairing state has not been reported.  A detailed description of a possible FFLO state in organic superconductors is beyond the scope of this article. For a recent review, see Ref.~[\citen{ReviewOrganics07}].

An FFLO state in layered high-$T_{\rm c}$ cuprates with strong superconducting fluctuations is highly unlikely, because the region of the FFLO state is replaced by a vortex liquid state below the mean field upper critical field line, which is a nonsuperconducting phase.

Besides heavy fermion and organic compounds, some Chevrel phase materias, like PbMo$_6$S$_8$ might be considered candidates for exhibiting the FFLO state~\cite{Dec82}, because of their very large upper critical field.  However, the very small mean free path of these materials ($\ell \ll \xi$) prevent the formation of the FFLO state. Borocarbide superconductor TmNi$_2$B$_2$C have also been suggested to be a promising candidate for the FFLO state because of a large paramagnetic effect~\cite{Hou01}.

\section{Possible FFLO state in heavy fermion superconductors}

We here address the issue of the superconducting states in CeRu$_2$, UPd$_2$Al$_3$, UBe$_{13}$, and CeCoIn$_5$.  In these materials, a possible existence of the FFLO state has been suggested by a number of groups.~\cite{Glo93,Yin93,Tac96,Buz97}   For the following discussion, the physical parameters  related to the FFLO transition in the normal and superconducting states of these materials are listed in Table I.

\begin{table*}[t]
\begin{center}

\caption{Physical parameters in the normal and superconducting states of CeRu$_2$, UPd$_2$Al$_3$, and CeCoIn$_5$.  The orbital limited upper critical fields in CeCoIn$_5$ are determined by eq.~(1).}

\footnotesize{}

\begin{tabular}{|c|c|c|c|c|c|c|c|c|c} \hline
                                                        & $T_{\rm c}$       &    $\ell$                           &    $\xi$   & $\kappa$ & $H_{\rm c2}$  & $\alpha$  & $\gamma$ &Gap symmetry    \\  
                                                        & (K)        &   (nm)                            &    (nm)   &  & (kOe) &  & (mJ/K$^2$mol) & \\  \hline

CeRu$_2$    &   6.1    &  130    &  7.9 &  25  &  52.3& 0.9 & 27 &anisotropic $s$-wave \\  \hline

UPd$_2$Al$_3$~~$\parallel a$    &   2.0    &   100   & 10.3 &  47   &  32  & 2.4~(polycrystal) & 145 &$d$-wave, $\cos(k_zc)$  \\  
~~~~~~~~~~~~~~$\parallel c$     &       &     &  9.3  &    & 38  &  &  & (horizontal node) \\  \hline

CeCoIn$_5$~~$\parallel a$   &  2.3   & $\sim 3000$     &  2.0 &  140   & 118~(Pauli), 386~(orbital)   & 4.6 & $\sim1000$ &$d_{x^2-y^2}$  \\  
~~~~~~~~~~~~~~$\parallel c$  &      &    & 4.3  & 44  &49.5~(Pauli), 177~(orbital)  & 5.0 &  &  \\  \hline

\end{tabular}

\normalsize{}

\end{center}
\end{table*}

\subsection{CeRu$_2$ and UPd$_2$Al$_3$: is the peak effect a signature of the FFLO state?}

CeRu$_2$ is a superconducting cubic Laves phase compound with $T_{\rm c} = 6.1\,$K. Since its discovery in 1958, CeRu$_2$ had been believed to be  a conventional $s$-wave type-II superconductor.  Recent experiments have revealed that the superconducting gap function is modulated at the Fermi surface, $\Delta_{\rm min}/\Delta_{\rm max} \approx 0.4$, indicating an anisotropic $s$-wave.~\cite{Kis05,Sak06}

UPd$_2$Al$_3$ is a heavy fermion superconductor ~\cite{Gei91} displaying quite unique properties, including the coexistence of  superconductivity and antiferromagnetic ordering with an atomic size local moment ($\mu=0.85\mu_{\rm B}$), and strong coupling between the magnetic excitation and superconductivity.~\cite{Sat01,Jou99}  Several experiments have revealed that the superconducting gap symmetry is most likely to be $d$-wave with line nodes parallel to the layers, $\Delta(\vk)=\Delta \cos k_zc$~\cite{Ber00,Wat04a,Mat06}.

These materials have attracted interest because a pronounced hysteresis behavior has been observed in a narrow field range below $H_{\rm c2}$, $H_{\rm i}<H<H_{\rm c2}$, in several different physical properties, such as the magnetization, ac susceptibility, ultrasound velocities, and thermal expansion.   On the other hand, these quantities are reversible in a wide field range below $H_{\rm i}$ down to the lower critical field $H_{\rm c1}$.  Particularly, at $H_{\rm i}<H<H_{\rm c2}$, the isothermal dc-magnetization  process in CeRu$_2$ and UPd$_2$Al$_3$ exhibits a hysteresis loop with a sharp peak at $H_{\rm peak}$ with decreasing $H$ and a sharp negative peak at the same field with increasing $H$. This peak effect arises from the critical state, indicating the occurrence of a peak in the critical current density of the system.

The origin of the peak effect has been discussed in relation to a possible FFLO state, since the vortex lines can become segmented into coupled pieces of lengths comparable to $\xi$, and become flexible in a qualitatively similar way as a pancake vortex in high-$T_{\rm c}$ cuprates\cite{Cle91} near the upper critical field.  These segments could conform more easily to pinning centers, thereby enhancing pinning.  In addition,  the transition line $(T_{\rm peak},H_{\rm peak})$ in CeRu$_2$ and UPd$_2$Al$_3$ resembles the FFLO transition line in the $H$--$T$ plane (see Fig.~\ref{fig:gg}, for example).  These features are also taken as possible signatures of the FFLO state. Moreover, the large electron mass and anisotropic gap symmetry of UPd$_2$Al$_3$ are favorable for FFLO formation.

However, subsequent studies have cast  doubt on the interpretation of the peak effect in these two compounds in terms of the FFLO transition.   For instance, the Maki parameter of CeRu$_2$ is less than unity, presumably because CeRu$_2$ is not a heavy fermion material and hence the orbital critical field $H_{\rm c2}^{\rm orb}$ is not large (see Table II). Moreover, the transition line $(T_{\rm peak},H_{\rm peak})$ splits from the $H_{\rm c2}$-line at $T_{\rm peak}/T_{\rm c} \approx 0.85$ for CeRu$_2$ and $\approx 0.80$ for UPd$_2$Al$_3$. These values are much larger than the theoretical predictions in Refs.~[\citen{Ful64,Lar64,Gru66}]. A possible explanation for the large value of $T_{\rm peak}/T_{\rm c}$ in UPd$_2$Al$_3$ is that the actual $T_{\rm c}$ at zero field is depressed by the pair-breaking effect due to the exchange field in the antiferromagnetically ordered state~\cite{Glo93}.  However, as pointed out in Ref.~[\citen{Nor93}], the calculation using the parameters obtained for UPd$_2$Al$_3$ provide a $T^{\dagger}$ which is much smaller than 0.56$T_{\rm c}$. Thus the features of the peak effect in CeRu$_2$ and UPd$_2$Al$_3$ are clearly at odds with the FFLO transition.

What then is the origin of the peak effect in CeRu$_2$ and UPd$_2$Al$_3$?   In the last decade, understanding of the peak effect has progressed. The peak effect is observed in clean superconductors with weak pinning.  Among them, the very anisotropic high-$T_{\rm c}$ cuprate Bi$_2$Sr$_2$CaCu$_2$O$_{8+\delta}$ and 2$H$-NbSe$_2$ have been widely studied due to their weak pinning properties.   It is widely accepted that in these systems the peak effect is ascribed to a disorder-driven first-order transition from the ordered Bragg-glass to the disordered vortex glass, caused by competition between the elastic and pinning energies.~\cite{Vin98,Gai00,Avr01}

An important question is: Is the peak effect in CeRu$_2$ and UPd$_2$Al$_3$ of the same origin as those in other clean superconductors, i.e., a disorder-driven transition generic to  weakly pinned vortices?   According to  subsequent studies of ac-susceptibility in 2$H$-NbSe$_2$ and CeRu$_2$, both materials display remarkable similarities in the peak effect region, which can be  explained by disorder-induced  fracturing and entanglement~\cite{Ban98} or transitions from collective to single-vortex pinning of the vortex lattice~\cite{Tro99}.  Based on these results, it has been concluded that the FFLO state in CeRu$_2$ is not required to explain the peak effect.  

Thus, subsequent research has cast doubt on the interpretation of the peak effect in CeRu$_2$ and UPd$_2$Al$_3$ in terms of the FFLO state.  At the present stage of  study, it is safe  to conclude that the FFLO state is very unlikely in CeRu$_2$ and UPd$_2$Al$_3$, though the detailed quasiparticle structure in the peak effect region should be scrutinized.

The observation of a possible FFLO state  has also been argued for UBe$_{13}$\cite{Buz97}, which exhibits a strong upturn in the upper critical field at low temperatures.~\cite{Ott84}  However,   recent NMR Knight shift measurements have revealed that the parity of UBe$_{13}$ appears to be spin-triplet with no Pauli paramagnetic effect on the superconductivity~\cite{Tou06}.

\section{Evidence for an FFLO state in CeCoIn$_5$}

\subsection{CeCoIn$_5$}

The recently discovered superconductor CeCoIn$_5$ is a very strong candidate for exhibiting the FFLO state.  CeCoIn$_5$ is a heavy fermion superconductor with a $T_{\rm c}$ of 2.3\,K, the highest among  Ce- and U- based heavy fermion superconductors.~\cite{Pet01}   It exhibits a layered structure of alternating conducting CeIn$_3$ planes (CeIn$_3$ is a superconductor under pressure) and less conducting CoIn$_2$ planes, suggesting a quasi-2D electronic nature.  This is supported by de Haas--van Alphen studies which reveal quasi-cylindrical sheets of the Fermi surface~\cite{Set01} and by NMR relaxation rate measurements which reveal the presence of quasi-2D spin fluctuation~\cite{Kaw03}.  Related to the quasi-two-dimensionality, the upper critical field for $\vec{H} \parallel ab$, $H_{\rm c2}^{\parallel}$, is nearly twice as large as that for $\vec{H} \perp ab$, $H_{\rm c2}^{\perp}$.  According to several experiments, including studies of the specific heat~\cite{Mov01}, thermal conductivity~\cite{Mov01,Iza01,Kas05}, NMR~\cite{Cur01,Koh01,Kaw03}, penetration depth~\cite{PEN}, and Andreev reflection~\cite{AND}, the superconducting gap symmetry of CeCoIn$_5$ is most likely to be $d_{x^2-y^2}$-wave with line nodes perpendicular to the plane.~\cite{Iza01,Aok04,Vek06}

The $H$--$T$ phase diagram of CeCoIn$_5$ exhibits several fascinating properties, which have not been observed in other superconductors. In a strong magnetic field, the superconducting transition is of first order at low temperatures for both configurations,  $\vec{H} \parallel ab$ and $\vec{H} \perp ab$, as shown in a number of experiments, including  jumps of the thermal conductivity\cite{Iza01,Cap04}, magnetization\cite{Tay02}, penetration depth\cite{Mar05}, ultrasound velocity\cite{Wat04}, thermal expansion\cite{Oes03}, magnetostriction~\cite{Cor06}, and NMR Knight shift~\cite{Kak05,Kum06}, and a sharp specific heat anomaly~\cite{Bia02,Mic06}. For $\vec{H} \perp ab$ and $\vec{H} \parallel ab$, the first order transition terminates at $\sim 0.3\,T_{\rm c}$ and   $\sim 0.4\,T_{\rm c}$, respectively, above which the transition becomes second order up to $T_{\rm c}$.  

The first order transition at the upper critical field indicates that the Pauli paramagnetic effect dominates over the orbital effect in both field directions.  According to Ref.~[\citen{Ada03}], for observation of the first order transition,  the system must be very clean.    Quantitatively,  for $\alpha=1$,  the transition becomes first order only when $\ell$ is more than 10 times longer than $\xi$.  These requirements are well satisfied in CeCoIn$_5$.  Indeed, the orbital limiting fields obtained from eq.~(\ref{eq:Hc2orb}) are 38.6\,T and 17.7\,T for $\vec{H} \parallel ab$ and $\vec{H} \perp ab$, respectively.  Then, assuming that the upper critical fields are close to the Pauli limiting fileds, $H_{\rm c2}^{\rm P} \simeq$ 11.8~T and 4.95~T for $\vec{H} \parallel ab$ and $\vec{H} \perp ab$, respectively, the Maki parameters determined by eq.\ref{eq:alpha} are estimated to be $\alpha^{\parallel} \simeq 4.6$ for $\vec{H} \parallel ab$ and $\alpha^{\perp} = 5.0$ for $\vec{H} \perp ab$.~\cite{Kum06}  Moreover,   microwave and thermal Hall conductivity measurements demonstrate that $\ell$ is more than 100 times longer than $\xi$ in the superconducting state, indicating that CeCoIn$_5$ is in an extremely clean regime.~\cite{Kas05,Deb06}

\subsection{FFLO phase in CeCoIn$_5$}

For $\vec{H} \parallel ab$, a second anomaly in the heat capacity has been observed within the superconducting state; a second order transition takes place in the vicinity of  $H_{\rm c2}^{\parallel}$ at low temperatures.~\cite{Rad03,Bia03}  This second-order transition line $(T^*_{\parallel}, H_{\parallel}^*) $ branches from the first order $H_{\rm c2}^{\parallel}$ line and decreases rapidly with decreasing $T$, indicating the presence of a new superconducting phase.   The second order anomaly $(T^*_{\parallel}, H_{\parallel}^*) $ has been confirmed by the subsequent studies of penetration depth\cite{Mar05}, thermal conductivity\cite{Cap04}, ultrasound velocity\cite{Wat04}, magnetization\cite{Gra06}, magnetostriction\cite{Cor06} and NMR\cite{Kak05,Mit06}.  This novel high field superconducting state at the low-$T$/high-$H$ corner of the $H$--$T$ plane has been identified as the FFLO state.

For $\vec{H} \perp ab$,  heat capacity measurements have also identified an anomaly  just below $H^{\perp}_{\rm c2}$.~\cite{Bia03} However, the study of the superconducting state has focused on the case of $\vec{H} \parallel ab$.   Recently, a new superconducting phase for $\vec{H} \perp ab$  has been confirmed by NMR\cite{Kum06} and magnetization\cite{Gra06} measurements.  From the analogy of the results for $\vec{H} \parallel ab$, the presence of  FFLO state for  $\vec{H} \perp ab$ has been suggested.

\begin{figure}[t]
\begin{center}
\includegraphics[width=8.0cm]{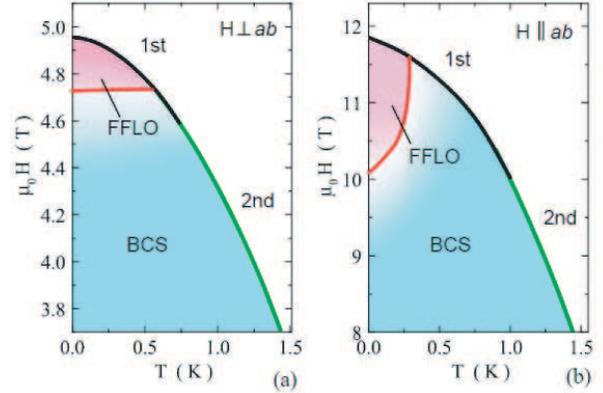}
\caption{Experimentally determined $H$--$T$ phase diagrams of CeCoIn$_5$ at low temperatures and high field for  (a)  $\vec{H} \perp ab$ and (b)  $\vec{H} \parallel ab$.  The colored portions display the FFLO (pink) and BCS (blue) regions.    The black and green lines represent the upper critical fields, which are first order and  second order, respectively.    The red lines represent the phase boundary separating the FFLO and BCS states. }
\label{fig:phasediagram}
\end{center}
\end{figure}

CeCoIn$_5$ appears to meet all the strict requirements placed on the existence of the FFLO state discussed in the beginning of \S3.  Figures~\ref{fig:phasediagram}(a) and (b) display the $H$--$T$ phase diagrams for $\vec{H} \parallel ab$ and for $\vec{H} \perp ab$, respectively, determined from several  experiments. The transition lines determined by different techniques are very close to each other. We next discuss the FFLO state as probed by various methods. 

\subsection{FFLO state in a parallel field}

\subsubsection{Penetration depth}

The penetration depth in the $ab$-plane for $\vec{H} \parallel ab$ has been measured at low temperatures near $H_{\rm c2}$.\cite{Mar05}   It has been shown that the penetration depth increases at $(T_{\parallel}^*,H_{\parallel}^*)$.  This can be interpreted as representing a decrease of the superfluid density due to the formation of the FFLO state.

\subsubsection{Magnetization and magnetostriction}

 Magnetization $M(H)$ as a function of $H$ has been measured by two groups near $H_{\rm c2}^{\parallel}$.\cite{Tay02,Gra06}  According to Ref.~[\citen{Gra06}], a change in the monotonic variation of the magnetization $M(H)$ is observed at $H_{\parallel}^* $.  This anomaly is clearly seen by plotting the field derivative of the magnetization $\d M/\d H$. The peak effect in the magnetization is not reported at $H_{\parallel}^*$, but is observed well below $H_{\parallel}^*$.\cite{Tay02,Gra06}

Magnetostriction measurements have also been performed as a function of $H$.\cite{Cor06}   The magnetostriction changes the slope at $H_{\parallel}^*$, indicating a second order transition.   Magnetostriction has also been measured in an oblique field.  The FFLO regime shrinks in an oblique field and when {\boldmath $H$} is tilted by 20$^{\circ}$ from the $ab$ plane, the FFLO regime occupies only  a very narrow regime below the upper critical field.

\subsubsection{Thermal conductivity}

Thermal conductivity is a unique transport quantity that does not vanish in the superconducting state.  It is particularly suitable for probing the quasiparticle structure, responding  only to  delocalized low energy quasiparticle excitations, since localized quasiparticles do not carry  heat.   The thermal conductivity is a directional probe of  quasiparticles.   In recent years, the thermal conductivity has been used effectively to probe the anisotropy of the order parameter in unconventional superconductors.\cite{Mat06}  This is due to the fact that normal delocalized quasiparticles are easily excited along the nodal directions.  In the FFLO state, anisotropy of the delocalized quasiparticle excitation spectrum is particularly noticeable, because the electron velocity vanishes along certain directions.  The inherent anisotropy of the quasiparticle excitation is expected to produce anisotropy of the thermal conductivity.

Thermal conductivity measurements in a parallel field have been carried out in two different configurations, $\vec{H} \parallel \vec{j}$ and $\vec{H} \perp \vec{j}$, near $H_{\rm c2}^{\parallel}$.\cite{Cap04}  Here $\vec{j}$ is the heat current.  The thermal conductivity  as a function of $H$ for $\vec{H} \parallel \vec{j}$ clearly displays a kink at $H_{\parallel}^*$  and exhibits an enhancement above $H_{\parallel}^*$.  In contrast, it is  difficult to  resolve these features for  $\vec{H} \perp \vec{j}$.

The observed enhancement of the thermal conductivity for $ \vec{H} \parallel \vec{j}$ is contrary to initial expectations, since in this geometry the FFLO nodal planes are created perpendicular to the heat current, and therefore  enhancement of the thermal conductivity would not be expected.   A possible scenario is the formation of a peculiar quasiparticle structure within the vortex core in CeCoIn$_5$, as will be discussed in \S 6.2.  Then, as suggested in Ref.~[\citen{Cap04}], the vortex core size might increase at the nodal planes, leading to enhancement of the thermal conductivity.  Further theoretical investigations into the thermal transport of the inhomogeneous FFLO state are necessary.

\subsubsection{Ultrasound velocity}

In order to study possible FFLO states, it is particularly important to determine the structure of the flux line lattice (FLL), which is intimately related to the electronic structure.    At the FFLO transition, the elastic modulus of the vortex lattice is expected to change, which can be detected by ultrasound measurements.   In ultrasound experiments, sound waves are coupled to the FLL when it is pinned by crystal lattice defects. The ensuing modified sound dispersion allows  detailed information about the FLL-crystal coupling to be extracted. More precisely, the change in the sound velocity with respect to the sound velocity in the normal state is~\cite{Dom96}
\begin{equation}
  \Delta v_{\rm t} = \frac{B^{2}}{\mu_{0} \rho_{\rm m} v_{\rm t}} 
    \frac{1}{1+(\Phi_{0}B)^{1/2}|k|^{2}/\mu_{0}j_{\rm c}},
\end{equation}
where $\rho_{\rm m}$ is the mass  density of the material, $k$ is the wave vector of the sound, and $j_{\rm c}$ is the critical current density.

\begin{figure}[t]
\begin{center}
\includegraphics[width=7.0cm]{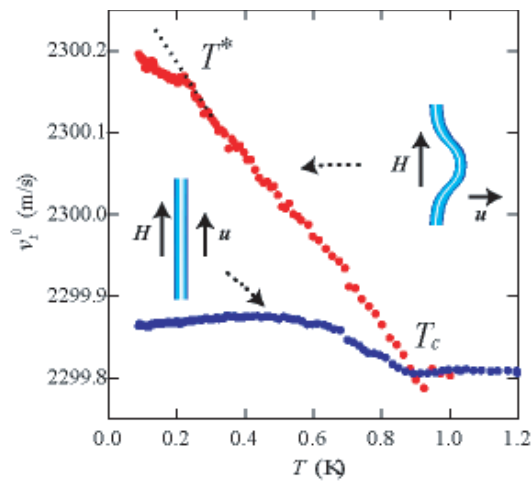}
\caption{Transverse sound velocity $v_{\rm t}^0$ for two different configurations, the Lorentz mode, $\vec{H} \parallel \vec{k} \parallel$ [100], $\vec{u} \parallel $[010],  and non-Lorentz mode,  $\vec{u}\parallel \vec{H} \parallel $ [100], $\vec{k}\parallel$[010].  In the non-Lorentz mode, the sound wave couples only to the crystal lattice, while the Lorentz mode corresponds to a flux line bending mode.   The difference between the two modes can be regarded as the response  from the flux line lattice.}
\label{fig:ultrasound}
\end{center}
\end{figure}

Figure~\ref{fig:ultrasound} displays the transverse sound velocity $v_{\rm t}^0$ for $\vec{H} \parallel ab$ as a function of temperature for two different configurations with respect to the polarization vector $\vec{u}$, the sound propagation vector $\vk$, and $\vec{H}$.\cite{Wat04}  We distinguish between the Lorentz force mode (Fig.~\ref{fig:ultrasound}, $\vec{u} \perp \vec{H}$),  where the sound wave couples to the vortices through the induced Lorentz force 
     $\vec{F}_{\rm L} \approx 
       \lambda^{-2} (\vec{u} \times \vec{H}) \times \vec{B}$, 
and the non-Lorentz mode ($\vec{u} \parallel \vec{H}$),  where the sound wave couples only to the crystal lattice. The transverse sound velocity in the Lorentz mode, in which the flux motion is parallel to the $ab$-plane,  is strongly enhanced compared to that in the non-Lorentz mode.  This indicates that the transverse ultrasound strongly couples to the FLL in the Lorentz mode.  The difference between the sound velocities in the Lorentz  and  non-Lorentz modes, 
     $\Delta v_{\rm t} =   v_{\rm t}^{0} (\vec{u} \perp     \vec{H}) 
                     - v_{\rm t}^{0} (\vec{u} \parallel \vec{H})$, 
can be regarded as being the contribution of the flux line lattice to the sound velocity, which provides  direct information of $c_{44}$ of the FLL.

 The enhancement of $v_{\rm t}^0$ in the non-Lorentz mode indicates a stiffening of the crystal lattice at the superconducting transition below $T_{\rm c}(H)$.  The enhancement of the sound velocity in the Lorentz mode is much greater than that in the non-Lorentz mode.  This immediately indicates that as the temperature is lowered the FLL becomes pinned.  In the Lorentz mode, the ultrasound velocity changes its slope with a cusp at $T_{\parallel}^*$.   It should be emphasized that such a cusp structure is not observed in the non-Lorentz mode.  This provides direct evidence that the transition at $T_{\parallel}^*$ is characterized by a change in flux line pinning. The sound velocity below $T_{\parallel}^*$ is smaller than that extrapolated from above $T_{\parallel}^*$, indicating that pinning is weaker below the FFLO transition.

Since the volume pinning force $F_{\rm p} = j_{\rm c}B$ is determined by the balance of the energy gain obtained by placing a flux line on a crystal defect, and the energy loss due to the ensuing deformation of the FLL, a decrease of pinning has two possible origins. Namely,  the decrease of the elementary pinning interaction between the defects and the flux lines, and the increase of one or several of the FLL elastic moduli. In the present case, the relevant elastic modulus is the dispersive flux line tilt modulus $c_{44}$. While a full theory is as yet unavailable, it is expected that the coupling of the deformations of the FLL and the nodal planes will lead to a slight decrease of $c_{44}$ in the FFLO phase~\cite{Ike06}. This would, however, lead to stronger pinning, as the FLL would be able to better adapt to the pinning impurities. The relatively smaller sound velocity in the FFLO state suggests that the first effect, a decrease of the elementary pinning interaction due to depression of the average order parameter magnitude, as indicated by the penetration depth and magnetization measurements, is the main effect.

Thus  ultrasound velocity measurements reveal an important decrease of the superconducting volume fraction at the FFLO transition. The nature and magnitude of the ultrasound velocity anomaly is quantitatively consistent with the modulated order parameter predicted for the FFLO phase.\cite{Wat04,Ike06}

\subsubsection{Nuclear magnetic resonance}

One of the most powerful probes for extracting microscopic information on quasiparticle and spin excitations is  NMR.    Here we discuss the NMR results in accordance with Ref.~[\citen{Kak05}], in which a dramatic change of the NMR spectra at $(T^*_{\parallel}, H^*_{\parallel}$) has been reported.    

\begin{figure}[t]
\begin{center}
\includegraphics[width=5.5cm]{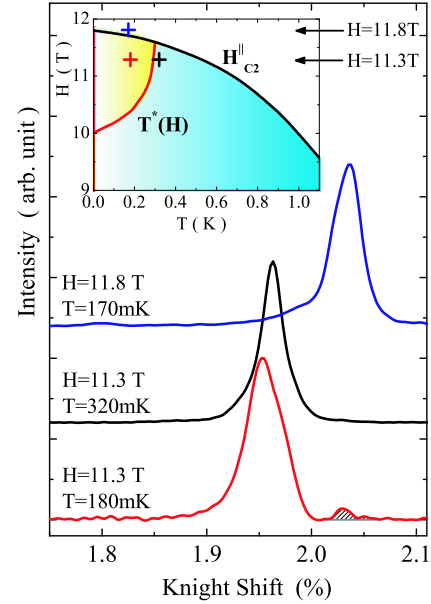}
\caption{Main panel: $^{115}$In-NMR spectra outside, slightly above $H_{\rm c2}^{\parallel}$ (blue  line),  slightly above $T^{*}$ (black), and well inside (red)  the FFLO phase.   The resonance feature at higher frequency is marked by hatching.  $(H,T)$-points at which each NMR spectrum was measured are shown by  crosses in the inset. Inset: Experimental $H$--$T$ phase diagram for CeCoIn$_5$ below 1\,K  for $\vec{H} \parallel ab$.   The regions shown in yellow and blue depict the FFLO and BCS regimes, respectively.   Horizontal arrows indicate the magnetic fields at which the NMR spectra were measured (see also Fig.~\ref{fig:nmr2}(a)). }
\label{fig:nmr1}
\end{center}
\end{figure}

Figure~\ref{fig:nmr1} displays the NMR spectra at the In(1) site ($^{115}$In ($I=9/2$)) with axial symmetry in the CeIn$_3$ layer, which is located in the center of the square lattice of Ce atoms.   (The tetragonal structure of CeCoIn$_5$ consists of alternating layers of CeIn$_3$ and CoIn$_2$ and hence has two inequivalent In sites per unit cell.)  The $^{115}$In NMR spectrum is plotted as a function of Knight shift $^{115}K$, which is a local spin susceptibility at the In(1)-site,  obtained from the central $^{115}$In line ($\pm 1/2 \leftrightarrow \mp 1/2$ transition).  For comparison, the spectrum in the normal state slightly above $H_{\rm c2}^{\parallel}$ (blue line), in the BCS state slightly above $T_{\rm c}^{\parallel}$ (black line), and well inside the FFLO state (red line), are shown in Fig.~\ref{fig:nmr1}.

A remarkable feature is the appearance of a new resonance peak with small but finite intensity at higher frequency in the spectrum well inside the FFLO phase, as clearly seen at $^{115}K\simeq2.03$\% (hatched region).  It is notable that the Knight shift of this new  higher resonance peak coincides well with that of the resonance line in the normal state above $H_{\rm c2}^{\parallel}$ (blue),  while the position of the lower resonance line is located close to that of the BCS state above $T^*_{\parallel}(H)$ (black).   Therefore, it is natural to deduce that the  higher resonance line originates from a normal quasiparticle regime, which  forms below $T^*_{\parallel}(H)$.   The appearance of the higher new resonance line, which is a manifestation of a normal quasiparticle structure, is  what is exactly expected for the FFLO state.  It is reported that the intensity of the higher resonance line is about 3--5 percent of the total intensity and is nearly $T$-independent below 180\,mK, indicating that the total volume of the planar node occupies  a few percent of the total volume.\cite{Kak05}

The well separation of the two resonance lines definitely indicates that the quasiparticles excited around the FFLO planar nodes are spatially well-separated from those excited around vortex cores.  This is again consistent with the topological structure of the vortex in the FFLO state, as discussed in \S 2.5, i.e., the quasiparticle regions within the vortex line do not spatially overlap with those in the FFLO planar nodes. 

\begin{figure}[t]
\begin{center}
\includegraphics[width=9.0cm]{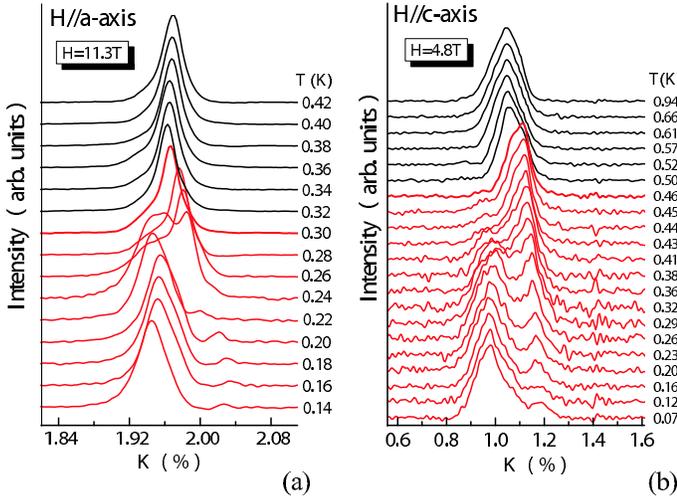}
\caption{$^{115}$In-NMR spectra as a function of Knight shift  from the normal state  (black lines) to the superconducting state (red lines) at (a) $H=11.3\,$T for  $\vec{H} \parallel ab$ (see the inset of Fig.~\ref{fig:nmr1}) and (b) $H=4.80\,$T for $\vec{H} \perp ab$.  The intensity is normalized by the largest peak intensity.  
}
\label{fig:nmr2}
\end{center}
\end{figure}

Figure~\ref{fig:nmr2}(a) displays the temperature evolution of the spectra at $H=11.3\,$T when crossing the $(T^*_{\parallel},H^*_{\parallel})$-line (see the horizontal arrow in the inset of Fig.~\ref{fig:nmr1}).     The higher resonance line grows rapidly with decreasing $T$ just below $T^{*}_{\parallel}(H)$.   A  double peak structure  appears at $T=240\,$mK,  followed by a shoulder structure at $T=260 \sim 300$~mK, indicating that the intensity of the higher resonance line dominates.   The two lines merge into a single line above $T^*_{\parallel}(H) \sim 300\,{\rm mK}$.

It has  been suggested that the temperature evolution of the NMR spectra  is compatible with the quasiparticle nodal planes formed in the FFLO state.\cite{Kak05}  The field induced layered structure expected in the FFLO phase resembles the superconducting states of stacks of superconductor--normal--superconductor  (S-N-S) Josephson tunnel junctions.\cite{Cle91,Mat95}  In  NMR experiments, {\it rf} shielding supercurrents flow across the FFLO planar nodes. Because of the second order transition at $T^*_{\parallel}(H)$,  the modulation length of the order parameter, i.e., the thickness of the superconducting layers, $\Lambda$ ($= 2 \pi/q$), diverges as $\Lambda \propto (T^*-T)^{-\alpha}$ with $\alpha>0$ upon approaching $T^*_{\parallel}(H)$.   Therefore $\Lambda$ exceeds the in-plane penetration length  in the vicinity of $T_{\parallel}^*$.  In this situation, an {\it rf} field penetrates into the normal sheets much deeper than into the SC sheets,  which results in a strong enhancement of the NMR intensity from the normal sheets.  At low temperature where $\Lambda$ becomes comparable to $\xi$ ($\ll \lambda$),  an {\it rf} field penetrates the normal sheets to the same degree as the superconducting sheets.  The temperature evolution of the higher resonance line can be explained semi-quantitatively by this model.\cite{Kak05}

To date, three groups have performed  NMR experiments to investigate the FFLO state in parallel fields.\cite{Kak05,Mit06,You06}   Although all groups have reported changes of the spectra in the vicinity of $H_{\rm c2}^{\parallel}$, the features of the spectra are different for each group and the interpretations of the spectra are controversial.  Ref.~[\citen{Mit06}] reports a change of the line width of the main spectrum at $(T^*_{\parallel}, H^*_{\parallel})$ without the appearance of the higher resonance peak.  In Ref.~[\citen{You06}],  NMR spectra  at the In(1) and  In(2) sites were measured in a wide frequency range.  Although no discernible change of the spectra at the In(1) site below $T_{\parallel}^*$ is observed, the appearance of a broad line at the In(2) site has been reported. This broadening is attributed to  possible antiferromagnetic (AFM) ordering within the  vortex cores, which will be discussed in \S5.6.2.   At present, the origin of the difference in the NMR spectra among the three groups is an open question and calls for further investigation.

\subsection{FFLO state in a perpendicular field}

Despite the observation of an anomaly in the heat capacity in a perpendicular field just below $H_{\rm c2}^{\perp}$,  the study of the superconducting state has focused on the $\vec{H} \parallel ab$ case.  Recent NMR and magnetization measurements provide several pieces of evidence for the FFLO state in the perpendicular field case as well.  NMR results in a perpendicular field have been reported by only one group.\cite{Kum06}  Figure~\ref{fig:nmr2}(b) depicts the temperature evolution of the NMR spectrum for $\vec{H} \perp ab$ just below  $H_{\rm c2}^{\perp}$ (4.95\,T at $T=0$). The spectrum and its temperature evolution bear  striking resemblance to those for $\vec{H} \parallel ab$ shown in Fig.~\ref{fig:nmr2}(b).   In the superconducting state,  the spectrum  exhibits a shoulder structure at lower frequencies ($T=0.46\,$K).    This shoulder structure is followed by a double peak structure as the temperature decreases,  clearly showing below $T=0.43\,$K.    The double peak structure persists down to the lowest temperature (70\,mK) and the intensity of the higher resonance line decreases with decreasing temperature.  As discussed in \S 5.3.5,  the appearance of the higher resonance peak is a manifestation of a new  quasiparticle region in the FFLO state.   Therefore, the NMR data for $\vec{H} \perp ab$  also leads  to the conclusion that the FFLO state is realized for $\vec{H} \perp ab$.

According to Ref.~[\citen{Kum06}], the double peak anomaly in the spectrum is not observed below  $H=4.7\,$T,  indicating that the critical field $H_{\perp}^*$ separating the FFLO and the BCS pairing state lies in a very narrow range between 4.7\,T and 4.75\,T.  Very recent magnetization measurements reports that $\d M/\d H$ exhibits a peak at $H_{\perp}^*$.  The results indicate that the phase boundary between FFLO and BCS states is nearly horizontal, i.e., $H_{\perp}^*$ is $T$-independent, which is consistent with the NMR results (see Fig.~\ref{fig:phasediagram}(a)).


Very recently, NMR spectra at the Co, In(1), and  In(2) sites have been measured in a wide frequency range for $\vec{H} \perp ab$ \cite{Oya07}. In contrast to the spectra for $\vec{H} \parallel ab$ reported in Ref.~[\citen{You06}],  in which a possible AFM ordering is suggested within the vortex cores as discussed in \S 5.3.5,  no anomalous broadening of the spectra has been observed in this configuration. Therefore,  there is no microsopic  evidence of the AFM ordering in the vortex core for $\vec{H} \perp ab$.

\subsection{$H$--$T$ phase diagram}

We here discuss the FFLO region in the experimentally determined $H$--$T$ phase diagram displayed in Figs.~\ref{fig:phasediagram}(a) and \ref{fig:phasediagram}(b).      The topology of the phase transition lines for both configurations $\vec{H} \parallel ab$ and $\vec{H} \perp ab$ is identical in that the FFLO line branches from the first order to normal transition line.   Despite the similarity of the phase diagrams of both configurations,  there are distinguishing differences between them.   For  $\vec{H} \perp ab$, the FFLO phase occupies a tiny high-$H$/low-$T$ corner in the $H$--$T$ phase diagram, in contrast to the case for $\vec{H} \parallel ab$.   In addition,  $H_{\perp}^*$ has almost no temperature dependence, in striking contrast to $\vec{H} \parallel ab$.

The appearance of the FFLO state in $\vec{H} \perp ab$ is not surprising, because as discussed in \S 5.1, the Maki parameter  $\alpha^{\perp}=5.0$  greatly exceeds the minimum value of 1.8 for the FFLO state.   We note that $\alpha^{\perp}$ is even larger than $\alpha^{\parallel}$.    The  extremely large $\alpha^{\perp}$ can be attributed to the strongly enhanced Pauli paramagnetic susceptibility in the normal state.   In fact,  the Pauli susceptibility for $\vec{H} \perp ab$ is nearly twice as large as that for $\vec{H} \parallel ab$.

The sizable decrease of the FFLO region for $\vec{H} \perp ab$ indicates that the FFLO state is more stable for $\vec{H} \parallel ab$.  This may be related to the quasi-2D Fermi surface of the system, as discussed in \S 2.3.  Recently, it has been proposed that the presence of a small sheet of 3D Fermi surface can stabilize the FFLO state in a perpendicular field.\cite{Ike06c}    In addition, the enhanced antiferromagnetic fluctuation in a perpendicular field compared with a parallel field may  be important for the stability of the FFLO state.

Some of FFLO theories assume a second order transition at the upper critical field, which is incompatible with the first order transition observed in CeCoIn$_5$.  The first order transition from the normal state to the FFLO state has been theoretically predicted  by several authors.~\cite{Mat98,Agt01,Bow02,Ada03,Dal04,Mor05,Hou06,Vor06}  .

Recently, the phase diagram in a quasi-2D system with $d$-wave symmetry has been examined  in the context of CeCoIn$_5$.~\cite{Ada03,Won04}  The second order FFLO-BCS transition that branches from the first order transition line at the upper critical field is reproduced in Ref.~[\citen{Ada03}], as shown in Fig.~\ref{fig:adachi}. However, an FFLO state with Landau level index $n=1$ at very low temperatures has not been reported experimentally.  Recently it has been pointed out that the FFLO state with $n=1$ is more unstable against the impurities than that with $n=0$.\cite{Ike06c}   For the second-order transition, the $H_{\rm c2}^{\parallel}$-curve has been semi-quantitatively reproduced.~\cite{Won04}  In addition, it has been pointed out that the fluctuation-driven first order transition by impurities is consistent with observations for CeCoIn$_5$.~\cite{Dal04,Ike05,Dal05}  The effect of the Fermi liquid correction has also been discussed in connection with CeCoIn$_5$.~\cite{Vor06}

\subsection{Another aspect of the FFLO state in CeCoIn$_5$}

\subsubsection{Influence of antiferromagnetic fluctuation}

The above results strongly suggest that an FFLO state exists in CeCoIn$_5$ for both parallel and perpendicular fields.  However, several unexpected features have been observed.  A most interesting and important feature is the competition or coexistence with AFM ordering.

The occurrence of superconductivity in CeCoIn$_5$ in the vicinity of the AFM quantum critical point (QCP) gives rise to pronounced non-Fermi-liquid behavior in various physical quantities in the normal state.\cite{Tay02,Bia03a,Nak04,Sid02,Kaw03}  Moreover, the specific heat and resistivity measurements have shown a crossover from Fermi liquid to non-Fermi liquid behavior in a magnetic field, which indicates the presence of a field-tuned AFM QCP at or in the vicinity of $H_{\rm c2}^{\perp}$.\cite{Bia03a,Pag03}  This implies that superconductivity persists, preventing the development of AFM order.  Moreover, it has been reported that AFM ordering appears when a magnetic field is applied in the superconducting state of CeRhIn$_5$, which exhibits superconductivity under pressure.\cite{Par06}

Then an important question is whether the origin of the the second order transitions at $(T_{\parallel}^*,H_{\parallel}^*)$ and $(T_{\perp}^*,H_{\perp}^*)$ are  related to some magnetic transitions.   Experimentally, the coexistence of an FFLO state and an AFM ordering in CeCoIn$_5$ has been a controversial issue, as discussed below.

For  $\vec{H} \parallel ab$,  as discussed in \S 5.3.5, recent measurements of NMR spectra report the appearance of a remarkable broad line at the In(2) site, which has been tentatively interpreted as an AFM ordering within the vortex cores\cite{You06}, as observed in high-$T_{\rm c}$ cuprates.\cite{Kak03,Lak01} The coexistance of an AFM ordering and an FFLO state is appeared to be possible, since an FFLO $\vec{q}$-vector is much smaller than an AFM wave number. On the other hand, there are several pieces of evidence that is against the AFM ordering in the high field superconducting phase of CeCoIn$_5$.  First, recent heat capacity measurements of CeCoIn$_5$ under pressure clearly demonstrate that the FFLO regime expands in the $H$--$T$ plane when the AFM fluctuations are suppressed by pressure.  This provides direct evidence that the AFM fluctuations are unfavorable for the stability of the FFLO state.  This is understood by considering that the FFLO state favors forming the paramagnetic regime, while the AFM fluctuations that prefer the singlet state prevent the formation of the paramagnetic regime.     Second,  under AFM order,  the alternating hyperfine fields would produce two inequivalent $^{115}$In(1) sites, which would give rise to two NMR resonance lines with equal intensities, but such resonance lines are not observed, as shown in Figs.~\ref{fig:nmr1}, \ref{fig:nmr2}(a), and \ref{fig:nmr2}(b).  Third,  an AFM transition generally leads to a large anomaly of the stiffness constant of the crystal lattice, which can be detected by ultrasound.  However, as shown in Fig.  ~\ref{fig:ultrasound}, no anomaly of the ultrasound is observed at $T^*_{\parallel}$ in the no-Lorentz mode.   Fourth, no signature of the transition line has been observed above $H_{\rm c2}^{\parallel}$, unlike the case of CeRhIn$_5$. These considerations imply that the AFM ordered moments induced by magnetic field in the vortex state for $\vec{H} \parallel ab$ are very tiny, if present.

For $\vec{H} \perp ab$,  no anomalous broadening of the NMR spectra has been observed as discussed in \S5.4\cite{Oya07}, in contrast to $\vec{H} \perp ab$.  In addition, the nearly $T$-independent horizontal FFLO--BCS boundary in the $H$--$T$ plane for $\vec{H} \perp ab$ shown in Fig.~\ref{fig:phasediagram}(a) is very different from the AFM order line reported in Ref.~[\citen{Par06}].   Therefore, at least for $\vec{H} \perp ab$, there is no evidence of the AFM ordering in the vortex state.

 Until now, there is no theory of what might be expected for the FFLO state if magnetism were induced by a magnetic field.  Further experimental and theoretical investigation is strongly required to clarify this issue. 


\subsubsection{Unusual vortex core}

It has been suggested that the NMR spectra also provide an important information on the nature of the vortex core, which in turn would be closely related to the FFLO state.   The main NMR spectra just below $H_{\rm c2}^{\parallel}$ (black and red lines in Fig.~\ref{fig:nmr1})   deviates significantly from that in the normal state above $H_{\rm c2}^{\parallel}$.    Since this regime is located at $\sim 1/3$ of the orbital limited upper critical field, a substantial portion ($\sim 1/3$) of the NMR intensity should originate from the vortex core.   Therefore a distinct deviation of the NMR spectra just below $H_{\rm c2}^{\parallel}$ from the normal state  indicates that the Knight shift within the vortex cores is significantly reduced from the normal value.  This  feature is in sharp contrast to that in conventional superconductors, where the Knight shift within cores coincides with that in the normal state.

As discussed in \S 5.3.5 and \S 5.6.1, a possible AFM vortex core has been suggested \cite{You06}, as observed in high-$T_{\rm c}$ cuprates.\cite{Kak03,Lak01}  This peculiar feature of the vortex core in CeCoIn$_5$ is also inferred from the  energy scale of the quasiparticle spectrum in the vortex core set by the confinement energy $\hbar \omega_0 \sim \Delta^2/\varepsilon_{\rm F}$~\cite{Kas05,Mat94a,Har94,Kum01}. For almost all superconductors, this energy level is of the order of $\mu{\rm K} - {\rm mK}$, which is negligibly small.  For CeCoIn$_5$, however, $\varepsilon_{\rm F}\sim$ 15 K and $\Delta\sim 5\,$K, so that  $\hbar \omega_0\sim 1.5\,$K and the vortex core spectrum becomes important at low temperatures,i.e., the system is in a quantum regime.  The small $\varepsilon_{\rm F}$ is due to the renormalized Fermi energy by the heavy electron mass.  Moreover, in CeCoIn$_5$,  $\ell \sim 3\,\mu {\rm m}$ and $\xi \sim 4\,{\rm nm}$, which  yields $\ell/\xi \sim 750$ at low fields, which is much larger than $\varepsilon_{\rm F}/\Delta \sim 3$.  This is in sharp contrast to other superconductors in which $\ell/\xi \ll \varepsilon_{\rm F}/\Delta$.  This implies that the vortex in CeCoIn$_5$ enters a new quantum regime, the superclean regime, that is difficult to access in most superconductors.\cite{Kas05} 


Although the possibility of achieving the superclean regime and AFM ordering within cores in CeCoIn$_5$   should be scrutinized, both effects are expected to affect the quasiparticle structure and phase diagram in the FFLO state. Further detailed investigations are necessary to clarify this issue.

\section{Summary}

Several heavy fermion and layered organic superconductors are considered good candidates for  exhibiting the FFLO state.  Particularly, the peak effect observed in the magnetization near $H_{\rm c2}$ has been attributed to the FFLO state in several heavy fermion superconductors.  However, subsequent studies have raised questions about the interpretation of the data.  Although a possible FFLO state has been suggested in several organic superconductors in parallel fields, no thermodynamic or microscopic evidence has been reported so far.  Of these superconductors, CeCoIn$_5$ appears to be a unique system for which corroborating experimental evidence supports the FFLO state.  In fact, CeCoIn$_5$ appears to meet all the strict requirements on the existence of the FFLO state.  The FFLO state in CeCoIn$_5$ is present for both parallel and perpendicular field directions.  The topology of the phase transitions in $H$--$T$ phase diagram for both configurations appear to be fully determined.  However, some of the microscopic measurements are still controversial and  some of the results seem to be different from those originally predicted for the FFLO state in some respects.

 For further clarification of the FFLO state, experiments using more direct techniques, such as neutron diffraction that can detect the spatial variation of the polarized spin excitations, and scanning tunneling microscopy that can directly detect the spatial variation of the quasiparticle density of states, are required.  Moreover  phase sensitive experiments, such as measuring the tunnel effect between the FFLO state and the conventional $s$-wave BCS state,  would provide an unambiguous method to detect the FFLO state.   For instance, it has been suggested that while the Josephson current between FFLO and  conventional BCS superconductors is suppressed, it can be recovered by applying a magnetic field~\cite{Yan00}. It has also been suggested that tunneling spectra between FFLO and conventional BCS superconductors have characteristic line shapes with several maxima and minima reflecting minigap structures due to the periodic pair potentials.~\cite{Tan06b}  Recently, it has been pointed out that the intrinsic pinning of vortices in the FFLO state can be used  as a direct probe of the spatial modulation of the FFLO state~\cite{Bul03}.

Although not discussed in this paper,  interesting states which resemble the FFLO state have been proposed for various systems.  For example, an FFLO-like state is induced near the interface between a ferromagnet and a superconductor.  In this case, the pair amplitude in the ferromagnetic region spatially oscillates  and as a result, a $0$-$\pi$ transition is expected. (For a review, see Ref.~[\citen{Bud05}].) In superconductors with no spatial inversion symmetry, two novel superconducting states are predicted, depending on the strength of the spin--orbit coupling $\lambda_{\rm so}$.   When $\lambda_{\rm so} \gg \Delta$,  a helical and a stripe vortex states have been proposed, in which the order parameters possesse a modulation $\e^{\i \vq \cdot \vr}$ \cite{Kau05} and $\cos({ \vq \cdot \vr})$ \cite{Agt06}, respectively,  which are similar to the FFLO state.    When $\lambda_{\rm so} \sim \Delta$, a possible triplet FFLO state induced by spin--orbit coupling in the absence of an external magnetic field has been proposed~\cite{Tan06c}.

\section*{Acknowledgments}

We thank  C.~Capan, R.~Ikeda, U.~Klein, K.~Kumagai, K.~Machida,  K.~Maki, K.~Nagai, D.~Rainer, and C.~J.~van der Beek for many clarifying discussions on the subject of this paper. We also thank H.~Adachi,   T.~Hanaguri, M.~Ichioka,  K.~Izawa, K.~Kakuyanagi, Y.~Kasahara, R.~Movshovich,  M.~Nohara, Y.~Onuki, T.~Sakakibara, T.~Shibauchi, S.~Shishido, M.~Tachiki, M.~Tanatar, Y.~Tanaka, S.~Uji, and T.~Watanabe for helpful discussions.  This work was partly supported by a Grant-in-Aid for Scientific Reserch from the Ministry of Education, Culture, Sports, Science and Technology.


\begin{thebibliography}{999}
\bibitem{OSC} 
  The QLRO is sustained in the states expressed 
  by eqs.~(\ref{eq:fflotri}), (\ref{eq:fflosqu}), and (\ref{eq:fflohex}). 
  See Ref.~[\citen{Shi98b}].
\bibitem{enhance}
  The Pauli limiting field $H_{\rm c2}^{\rm P}$ 
  is reduced by a factor $\sqrt{\chi_0/\chi}$, 
    while $H_{\rm c2}^{\rm FFLO}$ is reduced by $\chi_0/\chi$.
\end{thebibliography}
\end{document}